\documentclass[12pt]{article}
\usepackage{graphicx}
\usepackage{amstext}
\usepackage{float}
\usepackage{longtable}
\usepackage{cite}
\usepackage{authblk}
\usepackage{booktabs}

\title{Toward bio-inspired information processing with networks of nano-scale switching elements}

\author{Zoran Konkoli\thanks{zorank@chalmers.se}}
\author{G\"oran Wendin\thanks{goran.wendin@chalmers.se}}
\affil{Department of Microtechnology and Nanoscience\\ 
Chalmers University of Technology \\ 
SE-41296 Gothenburg \\
Sweden}

\begin{document}
\date{}
\maketitle
\begin{abstract}
Unconventional computing explores multi-scale platforms connecting molecular-scale devices into networks for the development of scalable neuromorphic architectures, often based on new materials and components with new functionalities. We review  some work investigating the functionalities of locally connected networks of different types of switching elements as computational substrates. In particular, we discuss reservoir computing with networks of nonlinear nanoscale components. In usual neuromorphic paradigms, the network synaptic weights are adjusted as a result of a training/learning process. In reservoir computing, the non-linear network acts as a dynamical system mixing and spreading the input signals over a large state space, and
only a  readout layer is trained. We illustrate the most important concepts with a few examples, featuring memristor networks with time-dependent and history dependent resistances.
\end{abstract}


\noindent{\it Keywords}: Reservoir computing, molecular network, memristor, dynamic system

\section{Background}

Petascale computers are now available \cite{Wikipedia_petaflop,Wikipedia_petascale_comp}, with a 2013 world record of 30 petaflops (Pflops) \cite{Tianhe-2}, and the roadmaps aim for exaflop computing by 2020 \cite{Wikipedia_exascale_comp,ASCAC2010,IESP,IESP-roadmap_2011,Dongarra2011}. These advances are made possible by using parallel processing (e.g. China''s Tianhe-2 Supercomputer \cite{Tianhe-2} uses 3 million cores). There are projections that zettaflop (1000 exaflop) computers are needed for long-range weather forecasting. Nevertheless, within a decade or two, the computational power of Turing-style digital high-performance  computing (HPC) is expected to level off for a number of reasons \cite{Sterling2011,Mooreslaw,Dennard1974,Haensch2006,Koomey2010,Moore2013}. This implies that in future we may lack the computational power needed to solve important problems that may be essential for further progress of mankind.
Several examples of important computational problems  have been presented in \cite{ASCAC2010}, e.g., understanding climate changes, reduction of the carbon footprint of the transportation sector, reverse engineering of the human brain, or design and manufacturing of advanced materials. 

The growth of computational power of Turing style digital computation following Moore's law \cite{Mooreslaw} is achieved in basically five ways: (1) scaling down the sizes of components, (2) scaling up the number of transistor in a processor, (3) scaling up the speed, (4) scaling up the number of processors (cores) on a chip, and (5) scaling up the number of multi-core processor boards in large (super)computers.

At the hardware level, computational power is given by the number of bit flips per second (related to Flops), which is then connected to electrical power since every bit is a transistor switching between "off" and "on" voltage (energy) levels. The stored energy has to be dissipated every clock cycle, which leads to local heating. 
The foundation for the impressive scaling described by Moore's law is Dennard's scaling law \cite{Dennard1974,Haensch2006}: transistor down-scaling is done in such a way that the electric fields in the channel are kept constant. When the linear scale of a transistor (channel) is reduced by a factor of 2, voltages and currents will be scaled down by a factor of 2. Consequently, the power and the area per transistor will scale down by a factor of 4. This implies that the {\em power density} will remain {\em constant}. 

The approaching end of Dennard scaling  \cite{Sterling2011,Moore2013} is due to the impossibility to scale down the voltages beyond limits set by reproducibility and robustness against static and dynamic fluctuations, i.e. inhomogeneities and noise.
Scaling up the transistor switching frequency will therefore increase the power density that has to be dissipated, and heating problems are presently limiting microprocessor clock frequencies to about 4GHz. Note that the problem is primarily not the heating of the transistors themselves, but the high-frequency charging of the capacitances of the metallic connection network, the energy of which has to be dissipated. To this comes heating due to leakage currents.

Nevertheless it seems that computational performance as a function of input power - computational (electrical) efficiency - has been following Moore scaling, implying that "The electrical efficiency of computation has doubled roughly every year and a half for more than six decades"  \cite{Koomey2010}. This is due to a range of optimisation efforts in microelectronics, including turning off parts of chips that do not have to be active at any given instant. As a result, a simple example is the increasing amount of computation one can get out of a single charging of a laptop battery. However, physical constraints must   limit also this type of scaling \cite{Koomey2010}, eventually  posing ultimate limitations on information processing power for solving complex or hard problems that require exponential growth of resources. On the road toward exa- and zetta-scale computing we will know what kind of paradigm shifts will be needed.

In this perspective, unconventional computing paradigms have been suggested to address and overcome a range of digital computing limitations (see e.g. \cite{UCOMP2009}).  There are several reasons why unconventional paradigms are needed, e.g., to investigate and road-map opportunities for  non-von Neumann computational paradigms, to solve exponentially hard problems by designing application specific devices, or to perform embedded computation in situations where large-scale CMOS solutions cannot be used.  

 As a specific example, an airplane flight-control system is basically an analogue system, but in modern implementations the control computer is simulated by a digital computer using analogue/digital interfaces. Following this line of thought toward complex systems and hard problems, analogue systems may be able to solve (optimise) problems that are hard for digital computers that might require exponential resources. 

This brings us to brain-inspired analogue neuromorphic approaches to computing, optimisation and control. Neuromorphic approaches emphasize the need for highly connected complex neural networks with adaptive synaptic-like connections. Since the 1980's, analogue VLSI for neural systems \cite{CarverMead1989} has been at the focus of strong efforts to build neuromorphic computers in semiconductor hardware. A current example is the Heidelberg CMOS-based large-scale hardware simulator \cite{Bruderle2011,Pfeil2012,Pfeil2013} implementing six networks on a universal neuromorphic computing substrate \cite{Pfeil2013}. Both for digital and analogue semiconductor nanoelectronics, by scaling down components one inevitably faces scalability problems with respect to charging/discharging, cross-talk, delays, losses and heating. The advantage of the analogue approach would then be, in applicable cases, that the computational power would be so great that the hardware does not need to invoke the most extreme CMOS or post-CMOS circuitry and processing speeds.

This has inspired visions about using self-organised networks of nanoscale components to mimic neural networks. The idea is that robust neuromorphic networks can be built from nano components with highly variable properties and erroneous behaviour.  In this review we will focus on this particular area of unconventional computing and describe efforts to use molecular electronics and nanoscale switching networks for digital and analogue/neuromorphic computing. This leads up to the recent promising field of reservoir computing, for which we will present some recent \cite{Kulkarni2012,KonkoliWendin2013} and new work on memristor-based networks.

\section{Unconventional computing}

The UCOMP 2009 consultation report \cite{UCOMP2009}, providing the background for a recent European FP7 call, states the following:
"Conventional digital Turing computation has been incredibly successful, but it encompasses only a small subset of all computational possibilities. Unconventional approaches to computation are currently less developed, but promise equally revolutionary results as they mature.
Unconventional computing is a broad domain, covering hypercomputation, quantum computing, optical computing, analogue computing, chemical computing, reaction-diffusion systems, molecular computing, biocomputing, embodied computing, Avogadro-scale and amorphous computing, self-assembling and self-organising computers, unconventional applications, and more."

In the long term, the vision is of course that unconventional schemes will be able to compete with post-CMOS high-performance computing by solving computationally hard problems that would take exponential time for Turing-like digital systems. However, in the shorter term, the goal is necessarily limited to exploring the computational capacity of unconventional systems without expecting to beat post-CMOS technology in what concerns speed and performance.

One unconventional approach is to explore the functionalities of locally connected switching networks. The computational substrate ("fabric") can then be a network of switches and memory devices, connected to input/output ports at the edges and embedded in a classical digital CMOS environment. Unconventional network components could be e.g. molecular transistors, negative differential resistances (NDR), or memristors. Although one often talks about "neuromorphic computing", it is important to realise that this is very far from brain-like behaviour. Artificial neural networks most often involve networks of synapses connected to simple integrate-and-fire neurons with much less functionality than biological neurons. Moreover, even if a molecular network might constitute a network of synapses, the neurons must in practice be built from external semiconductor circuits. 

Advanced 2-terminal characteristics most likely arise from various combinations of hysteretic switching, negative differential resistance (NDR) and rectifying behavior. Such strongly non-linear properties often emerge in nanoscale devices due changes in interfaces, dielectrics, or electronic structure under the influence of very strong biasing and gating fields, electric or magnetic. Recent examples are various kinds of memristors \cite{Alibart2009,Alibart2012}. At the same time, to produce such effects in reproducible ways in functional components is a great challenge, and the field is at an embryonic stage, especially in what concerns truly molecular devices. Nevertheless, 2-terminal components are of great interest because they are potentially scalable by just shrinking the junction area: there is no need for additional layers of gating contacts and wires, and there are realistic opportunities for self-assembly of 2D and stacked devices. Such properties are of great interest for future nanosystem development. In the following sections we will therefore describe the experimental status of devices of potential relevance for molecular scale nanoelectronics, and also discuss the theoretical progress. 

Of particular recent interest is reservoir computing with networks of nonlinear nanoscale components. In usual neuromorphic paradigms, the network synaptic weights become fixed as a result of a training/learning process. In conventional reservoir computing, however, one considers from the start a large network with a fixed set of weights. 
A set of input signals become mixed in the non-linear network, spreading over a large state space, and in a simple output layer (filter) the synaptic weights are trained to read out desired patterns. It seems extremely relevant and challenging to investigate the network dynamics resulting from multi-port dynamic signals, measure correlation functions, and analyse the possibility to read out specific information, e.g. using trained output networks as filters.

%
%

\section{Unconventional nanoelectronics}

The purpose of this section is to briefly review various nanoelectronic approaches aiming at meeting Moore's law issues with non-CMOS schemes for unconventional information processing. The intention is to motivate reservoir computing as the potentially most promising paradigm for self-organised nano/molecular electronics.

\subsection{Single-molecule devices and networks} 

The history and status of single-molecule electronics is extensively described in some recent reviews  \cite{Szaciowski2012,Prauzner2012,Soe2011}. 
A standard approach has been to design networks with single-molecule 2- and 3-terminal switches for digital logic and memory. Experimentally, such circuits have hardly been built, and can anyway not be expected to compete with CMOS. 

Of greater interest, and possibly more promising, are top-down approaches to connect single molecules, or small molecular clusters, to multiple electrodes for wave-function manipulation and quantum-state information processing  \cite{Joachim2012,Okawa2012}. Here the inputs are classical, but the logic gate is based on quantum-state interference, and STM (classical) readout. Okawa et al. \cite{Okawa2012} have demonstrated experimentally some first steps toward systematic construction of multi-terminal devices based on single phtalocyanine molecules, and Prasongkit et al. \cite{Prasongkit2011} have shown theoretically how a phtalocyanine-based interference-driven switch may work.

\subsection{Multi-molecule devices and networks} 

The dominating approach toward molecular electronics involves using self-assembled monolayers (SAM) between metal electrodes for building molecular junctions for switching and memory  \cite{Stewart2004,Green2007,Coskun2012}. In the original work of the HP-group  \cite{Stewart2004} it turned out that the switching effect was not caused by internal rotaxane molecular switching but rather by field-induced gold filaments. (Recently, however, both effects have been identified  \cite{Coskun2012}). The metal-molecule interface can thus be of paramount importance for the device functionality  \cite{Jia2013}. This type of (originally unwanted) interface effect is now at the heart of several types of memristors. 

\subsection{NDR devices and networks}

In 1999, Chen et al.  \cite{Chen1999} published an experimental paper reporting strong effects of negative differential resistance (NDR) in a molecular junction. This formed the basis for a number of theoretical papers using a network of NDR molecules (Nanocell) for implementing logic gates, with the particular advantage of producing XOR gates   \cite{Reed2001,Tour2002,Husband2003,Skoldberg2007a,Skoldberg2007b,Chiragwandi2010}. Unfortunately the original NDR result could not be reproduced, and strong molecular NDR effects have turned out to be difficult to produce, and so far Nanocell logic remains a theoretical issue. Note, however, that strong molecular NDR has recently been found experimentally \cite{Koole2013}. 

\subsection{Memristor devices and networks}

The memristor (memory resistor) was invented by Chua in 1971  \cite{Chua1971,ChuaKang1976} and further investigated and exploited by HP-researchers   \cite{Strukov2008,Yang2008,Borghetti2009}. The characteristic property of the memristor is that the resistance depends on a time-dependent state variable $x(t)$ and is itself time dependent, $R(x,V,t)$, resistance changes being controlled/switched e.g. by voltages that exceed certain thresholds. This type of component actually already had a long history (see \cite{WaserAono2007} for discussion and references), but the recent interest has lead to an explosive development  \cite{Alibart2009,Alibart2012,Jo2010,Chang2011,PershinDiVentra2010,PershinDiVentra2011,DiVentraPershin2012,PershinDiVentra2012,Bichler2012,Bichler2013} (see  \cite{Thomas2013,Yang2013} for recent overviews and more references). In particular, the function of the memristor depends on its history, which can be used for implementation of conditional logic  \cite{Borghetti2009}. Memristors show properties similar to biological synapses \cite{Alibart2012}, which means that they can be used for implementing and training neuromorphic networks. A well-known example is that of Pavlov's Dog \cite{PershinDiVentra2010,Bichler2013}: here two separate time signals are input on the Food and Bell terminals of a Perceptron. Initially the weight of the Bell synapse is low, and the Bell signal alone has no effect. However, via feedback the Food and Bell signals can act together over the Bell synapse (memristor) and exceed the voltage threshold for changing and increasing the weight of the Bell synapse. In the end, the output will spike in response to the Bell signal alone. This effect was demonstrated experimentally (i.e. not simulated) with NOMFET organic synaptic transistors  \cite{Bichler2013}.

%

\section{Reservoir computing (RC)}

\begin{figure}[H]
\begin{flushright}
\includegraphics[width=0.9\columnwidth]{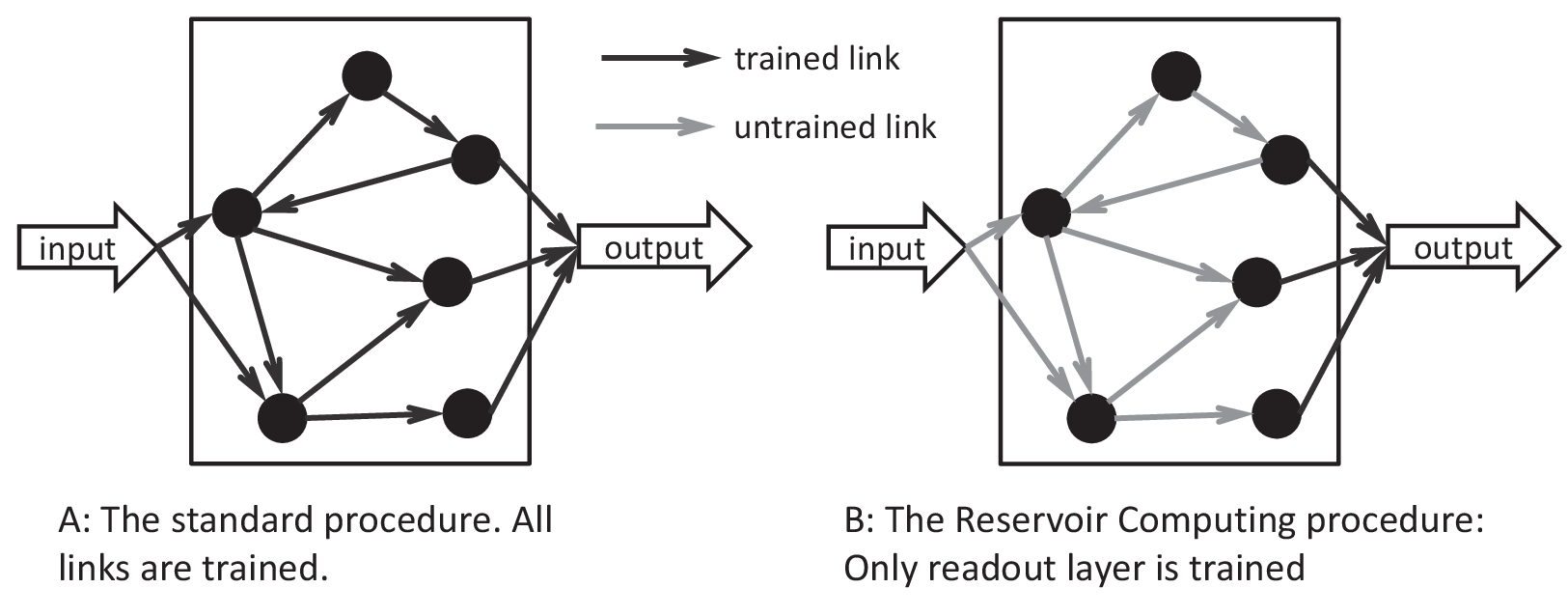}
\end{flushright}
\caption{\label{fig:figRCIdea}
The main idea behind Reservoir Computing. A simple readout mechanism (single-layer synaptic filter) is trained to read the state and map it to the desired output. The training is performed only at the readout stage and the reservoir is fixed in principle. In such a way there is no need to train large networks.  (Inspired by~\cite{Lukosevicius2012}.) }
\end{figure}

Of particular recent interest is reservoir computing with recurrent neural networks (RNN) (involving feedback) of nonlinear nanoscale components. In usual neuromorphic paradigms, the network synaptic weights become fixed as a result of a training/learning process. In reservoir computing, however, one considers from the start a large network with a fixed set of weights and only trains a readout layer (as illustrated in Fig.~\ref{fig:figRCIdea}). The standard practise is to use linear readout layers. However,  it is possible to exploit more complicated readout layers provided they are easy to train. In what follows no formal distinction will be made between readout layer types. The term reservoir computing will be used to refer to any readout layer that is easy to train.

A reservoir computer \cite{RC2013,Jaeger2001,Natschlager2002,Maass2002,Natschlager2004,Bertschinger2004,Jaeger2004,Lukosevicius2009,Lukosevicius2012,Yildiz2012,Manjunath2013,Fernando2003,Ju2013,Busing2010,Yamazaki2007,Appeltant2011,Dambre2012,Paquot2012,Larger2012}  is a high-dimensional non-linear dynamical system, called reservoir, driven by time-dependent inputs. The dynamics map the input to a superposition of instantaneous internal states of the reservoir, carrying information about the input signals. In such a way, initial information contained in the input is spread into a space with many dimensions (states). The readout layer is used to pick a particular set of states.

Liquid-state machines (LSM) \cite{Natschlager2002,Maass2002,Natschlager2004,Bertschinger2004}, and echo state networks (ESN) \cite{Jaeger2001,Lukosevicius2009,Lukosevicius2012} represent two major types of reservoir computing (RC). The essence of RC is captured by a pedagogical "toy experiment"  \cite{Fernando2003}, implementing RC in a "bucket of water", presenting a real liquid state computer. Here the state of the machine is directly visible as an interference pattern that can be read out by an image processing system, e.g. a neural network. This means that the properties of a natural dynamical system can be harnessed to solve nonlinear pattern recognition problems and that a set of simple linear readout elements will suffice to make the classification \cite{Fernando2003}. Generically, this means that the state configuration generated by the input signals can be regarded as an internal interference (correlation) pattern that can be read out by a generic "image" processing device, typically a trained neural network. The spreading of the input signals over a large state space of the dynamical systems can be viewed as giving rise to a time-dependent pattern in state space, corresponding to dynamical patterns in real space (e.g. wave patterns), energy (frequency) and time.


Of particular interest for us is the very recent work by Kulkarni and Teuscher  \cite{Kulkarni2012} implementing RC in software for memristor-based networks with 5-40 nodes. The authors demonstrated two applications of memristor networks for information processing. In the first example a  readout layer of neurons (Perceptron) was trained to distinguish between sawtooth and square wave forms. 
 In the second example a version of the Pavlov's Dog  problem \cite{PershinDiVentra2010,Bichler2012} has been implemented. Two separate time signals are input on two sets of network terminals and the network correlations again read out by a readout single layer of synaptic elements. The output network is then able to learn to identify the Bell signal in the absence of the Food signal.
In the following we will illustrate the potential of using memristor networks in the context of reservoir computing by discussing the quality of the reservoir in a rather informal way.

\section{Building networks from memristive elements}



A memristor is a non-linear resistor element where time-dependent resistance \(R(x,V_M,t)\) is controlled by a time-dependent state function \(x(t)\), the rate of change of which is determined by a function \(f(x,V_M,t)\):
%
%
\begin{eqnarray}\label{HH}
V_M(t) = R(x,V_M,t) \; I(t)\\
\frac{dx(t)}{dt}  = f(x,V_M,t)
\end{eqnarray}

\begin{figure}[H]
\begin{flushright}
\includegraphics[width=0.9\columnwidth]{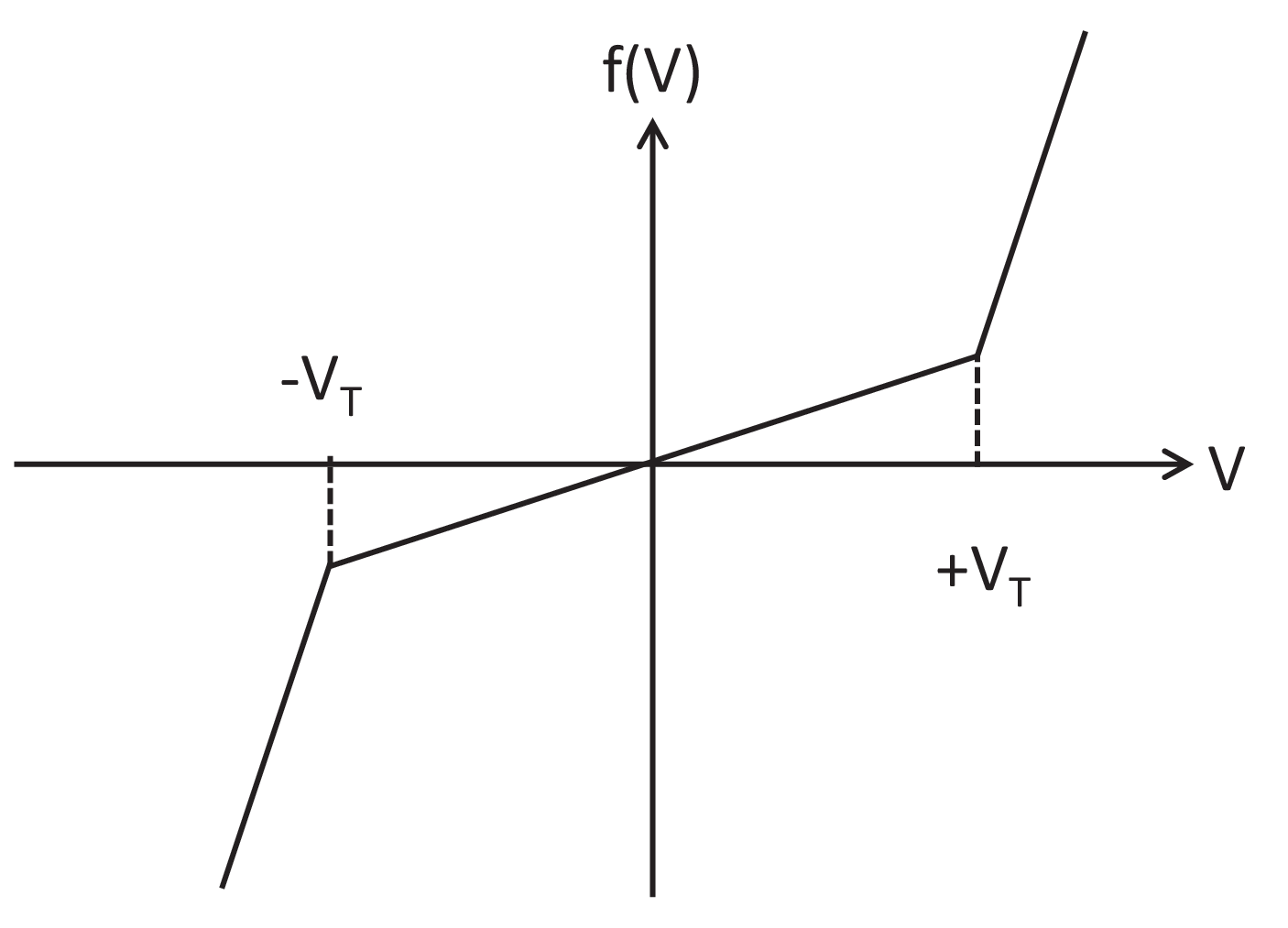}
\end{flushright}
\caption{\label{fig:figFuncF}
A typical dependence of the function \(f\) on the voltage \(V\equiv V_M\) applied to the memristive element. There is also an implicit dependence on \(R\). The depicted dependence on \(V_M\) holds only for  \(R_\text{min}\le R \le R_\text{max}\). Outside of this interval there is no change in the resistance, \(f(V_M)\equiv 0\).}
\end{figure}

A particularly interesting case is when the state function \(x=x(t)\) is switchable, slowly varying with time inside a low voltage range \(-V_T <V_M <+V_T\), but rapidly varying outside, as shown in Fig.~\ref{fig:figFuncF}. This standard memristor model~\cite{PershinDiVentra2010} is based on the following form of \(f\):
\begin{equation}
   f(V) = \beta V_M + 
       \frac{1}{2}(\alpha - \beta)  
        \left(\vert  V_{M} + V_{T}\vert - 
        \vert  V_{M} - V_{T}\vert
        \right)
\end{equation}

This functional form describes the majority of the experimental implementations of the memristor element. A common situation is that the state function can be represented by the resistance itself, \(x=R(t)\). The change of the resistance is typically induced by the bias voltage reversibly driving structural changes in the material, like field-induced diffusion of vacancies or creation of conducting filaments. Physical constraints then set limits on minimum and maximum resistance of the memristor. 

\begin{figure}[H]
\begin{flushright}
\includegraphics[width=0.9\columnwidth]{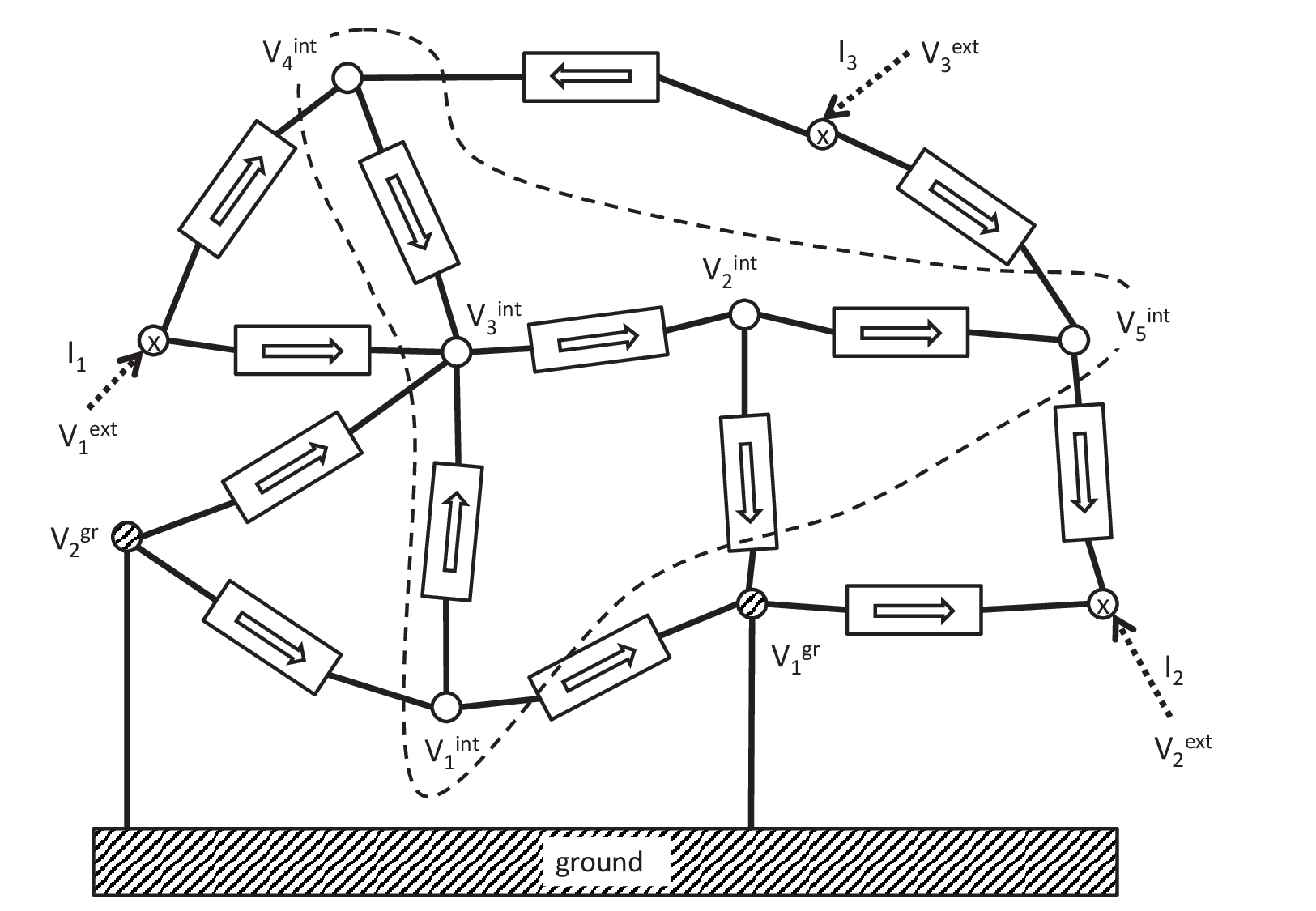}
\end{flushright}
\caption{\label{fig:figNetworkStructure}
An arbitrary memristor network (from \cite{KonkoliWendin2013}). The current is being supplied into the device through external nodes (\(V_{1}^\text{ext}\), \(V_{2}^\text{ext}\), \(V_{3}^\text{ext}\)) and drained through the ground contacts (\(V_{1}^\text{gr}\), \(V_{2}^\text{gr}\)). Voltages on the internal contacts (\(V_{1}^\text{int}\), \(V_{2}^\text{int}\), \(V_{3}^\text{int}\), \(V_{4}^\text{int}\), \(V_{5}^\text{int}\)) adjust according to the particular values of the resistance of the memristive elements. Note that each memristive element has a direction.
}
\end{figure}

   In the following we will discuss some typical features of memristor networks in the context of reservoir computing.
As an illustration, a simulation of a typical memristor network will be analyzed with the main goal to emphasize some key features of such networks. To set the stage, Fig.~\ref{fig:figNetworkStructure} shows how an arbitrary memristor network  is structured. The graph consists of nodes (contacts) that are connected by links (memristive elements).
Note that the memristor has a well-defined directionality since the function \(f(V)\) is an odd function of \(V\). This fact is indicated by the presence of the small arrow associated with each memristor element.

Surprisingly, there are not that many software packages that one can use to simulate memristor networks. We have developed a network software simulator in the Mathematica platform, the MEmristor NEtwork Simulator (MENES) package \cite{KonkoliWendin2013}. The MENES package accepts very generic input (network structure). The present code is structured in such a way that it can be easily implemented in any other computer language and modified toward various special purpose applications (e.g. it is relatively straightforward to implement other link types). This code will be used to perform the simulations below.

\begin{figure}[H]
\begin{flushright}
\includegraphics[width=0.45\columnwidth]{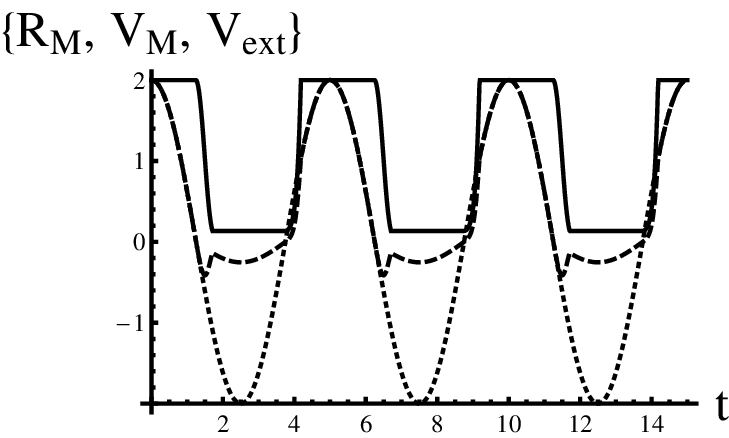}
\includegraphics[width=0.45\columnwidth]{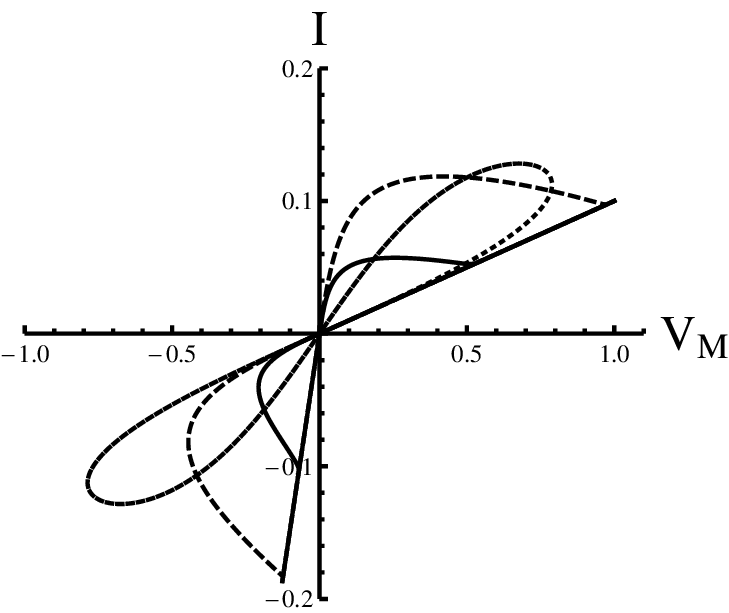}
\end{flushright}
\caption{\label{fig:benchmarking}
Simulation of the network with a passive resistor \(R=10k\Omega\) and a single memristor element connected in series. The memristor parameters are:   \(R_M^\text{min} = 675\Omega\);  \(R_M^\text{max} = 10k\Omega\); \(R_M(t=0)= 10k\Omega\); \(\alpha = \beta = 146k\Omega (Vs)^{-1}\); \(V_T = 4V\). The network is driven by \(V_\text{ext}(t)=V_0 \cos[2\pi\nu t]\) at several frequencies with \(V_0=2V\). The left panel: The memristance \(R_M(t)\) (the full line, in units of \(5k\Omega\)), the memristor voltage drop \(V_M(t)\) (dashed, in units of \(0.5 V\)), and the external drive \(V_\text{ext}(t)\) (dotted, in units of \(1V\)), depicted as a function of time \(t\) (in seconds) when the system is driven with \(\nu_1=0.2\text{Hz}\). The memristance plot shows typical hard limits of the resistance. The right panel: The parametric plots of \((V_M(t),I(t))\) for the three drives;  \(\nu_1=0.2\text{Hz}\) (the full line), \(\nu_2=1\text{Hz}\) (dashed), and \(\nu_3=5\text{Hz}\) (dotted). Again, note the effects of the hard limits (the linear parts corresponding to the plateaus of the resistance).}
\end{figure}

The elementary case of a single memristor is shown in Fig.~\ref{fig:benchmarking}.  Exactly the same curves were obtained by Pershin and DiVentra \cite{PershinDiVentra2010,PershinDiVentra2011} and Bichler \cite{Bichler2010}, serving as a benchmark in \cite{KonkoliWendin2013}. Note the typical hard limits on the resistance. If the external voltage stays within the \(\pm V_T\) limits the element behaves as a usual resistance. However, when the limit is exceeded non-linear effects start showing up. In this regime one expects non-trivial dynamics that can be exploited for information processing. In the next section we deliberately consider an example where the external voltages drive the memristance values so that their hard limits are frequently reached.

\section{An illustrative example: A self-assembled network}

\begin{figure}[H]
\begin{flushright}
\includegraphics[width=0.9\columnwidth]{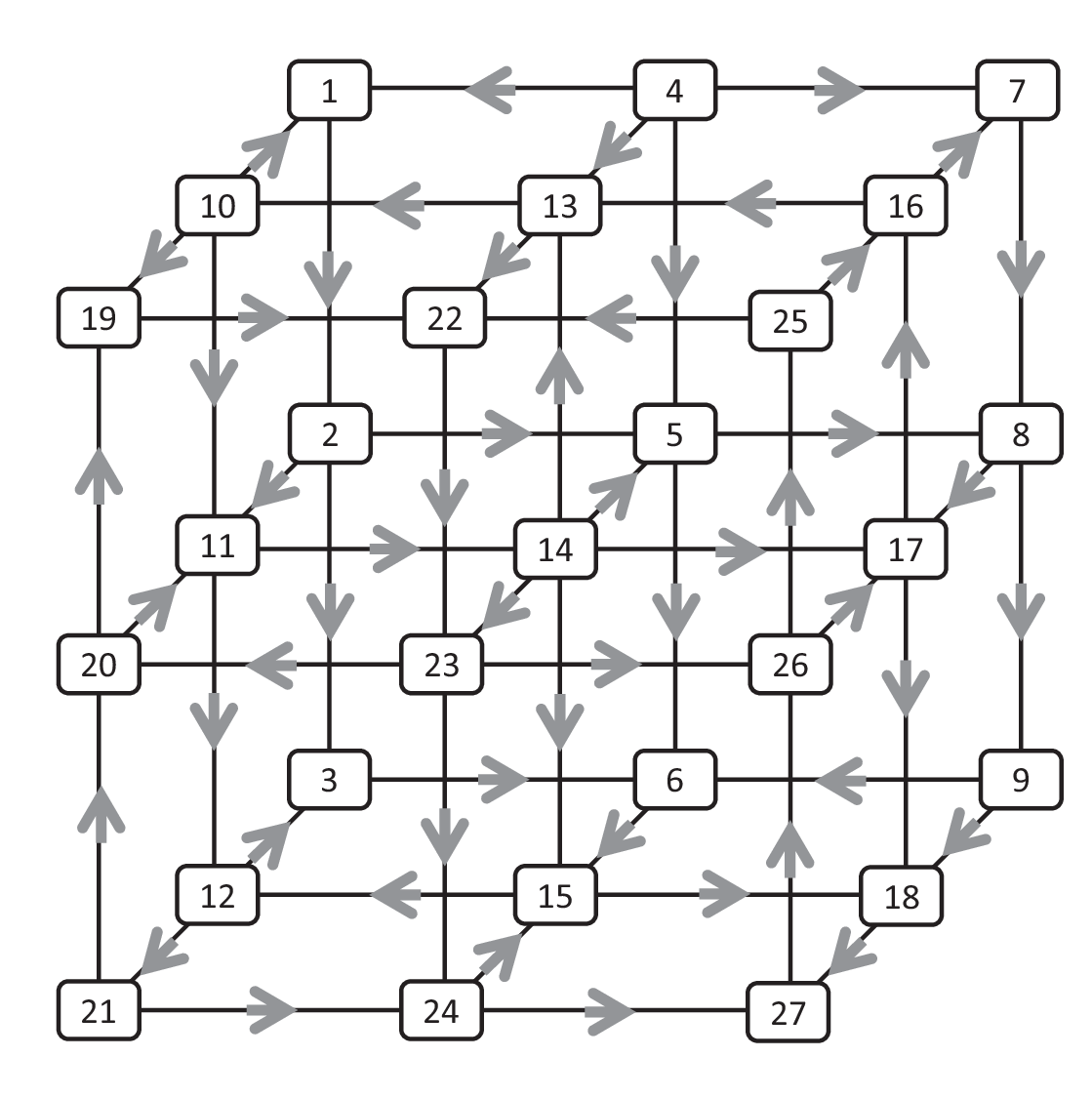}
\end{flushright}
\caption{\label{fig:figCube}
A memristor network arranged in a simple cubic 3x3x3 structure. There are  27 nodes in total where nodes 1, 2, and 3 are external nodes, node \(4\) is grounded, and the remaining 23 nodes are internal.    All parameters that define the memristor elements are detailed in appendix \ref{app:networkDetails}.
}
\end{figure}

Figure~\ref{fig:figCube} shows  a typical structure that is expected as a result from a generic nanotechnology assembly process where only local connections are present. The network consists of 27 nodes arranged in a simple cubic 3x3x3 structure. To illustrate the richness of the internal state space we show  how the device behaves under the influence of relatively simple external input.  It is very likely that the device can be addressed only through surface contacts. There are three input terminals (external nodes) and one grounded node (all at the surface of the device). The remaining 23 internal nodes may serve as output terminals. In practise, it would be very hard to address all contacts that are chosen as internal. Some of these contacts will occur deeply in the device interior. However, to simplify the discussion that follows, all internal contacts will be considered accessible.  There is a purpose behind  choosing the input nodes and the grounded node very close to each other: The goal is to further emphasize simplicity of the input by localizing all external contacts to a very narrow region of the device.  

Figure~\ref{fig:figCubeVextVintRM} shows the results of a simulation where the three periodic input signals   \(V_\text{ext}(t) = V_0 \sin(2 \pi \nu t)\) with equal amplitudes (\(V_0 = 1 \text{V}\)) were applied to the external nodes. The following frequencies were used:   \(\nu_1=2/T\) (applied at node 1), \(\nu_2=3/T\) (node 2), and \(\nu_3=5/T\) (node 3) with \(T = 1 \text{s}\). The implicit Euler method was used to integrate the system from \(t=0\) to \(t = 3T\) in \(N = 500\) steps (time increment \(dt=0.006\text{s}\)). 

A plain visual inspection of the figure suggests that the dynamics of the internal states is extremely complex.  If the system is to work in the context of reservoir computing, it is important that the input maps onto a rich set of internal states. The question then is
how similar or different are the intrinsic voltages relative to the external driving ones? In order to address this question we designed a relatively simple mathematical procedure to quantify the degree of distinction between a given output signal (an internal voltage or a memristance). 

The procedure works as follows: It is assumed that an output \(o(t)\) can be represented as a linear combination \(z(t)\) of time shifted input signals (in this particular case the external voltages). We find the best possible fit for the linear weights (they can be complex numbers) so that \(\Delta =||o(t)-z(t)||\) is the smallest possible, where \(\|.\|\) denotes a norm. The deviation  from the output signal and the best linear combination quantifies the degree of dissimilarity. For small values of \(\Delta\) the output signal can be reproduced as a linear combination of the input signals. However, for large values of \(\Delta\) the output signal is very different from the input. In such a case one expects that the input excites distinct internal modes of the system that are very different from the input. To carry out the procedure mathematically we work in the Fourier space instead and minimize \(\|o(\omega)-z(\omega)\|\) averaged over all \(\omega\). The dissimilarity measure is computed as 
\begin{equation}
   \delta = \frac{\|o(\omega)-z(\omega)\|}{\|o(\omega)\|}
\end{equation}
All computations were done in \textit{Mathematica} using the \textit{LeastSquares} and \textit{Fourier} functions provided by the package.

Figure \ref{fig:figCubeFourierSpectrumVint} shows the Fourier spectra \(FS(\omega)\) of  the output  voltage, \(o(\omega)\),  and the weighted superposition of input voltages, \(z(\omega)\). The first two panels depict the two \(o(\omega)\) voltages that are hardest to mimic by the weighted optimized linear combinations \(z(\omega)\) of input signals (their dissimilarity measures are largest). The last panel depicts the voltage with the smallest dissimilarity measure (that is easiest to mimic by the linear combination of inputs). Note that there is nevertheless a very good match in all cases. This implies that in general the internal node voltages only propagate the information contained in the input. This is an example of a poor reservoir. 

\begin{figure}[H]
\begin{flushright}
\includegraphics[width=0.7\columnwidth]{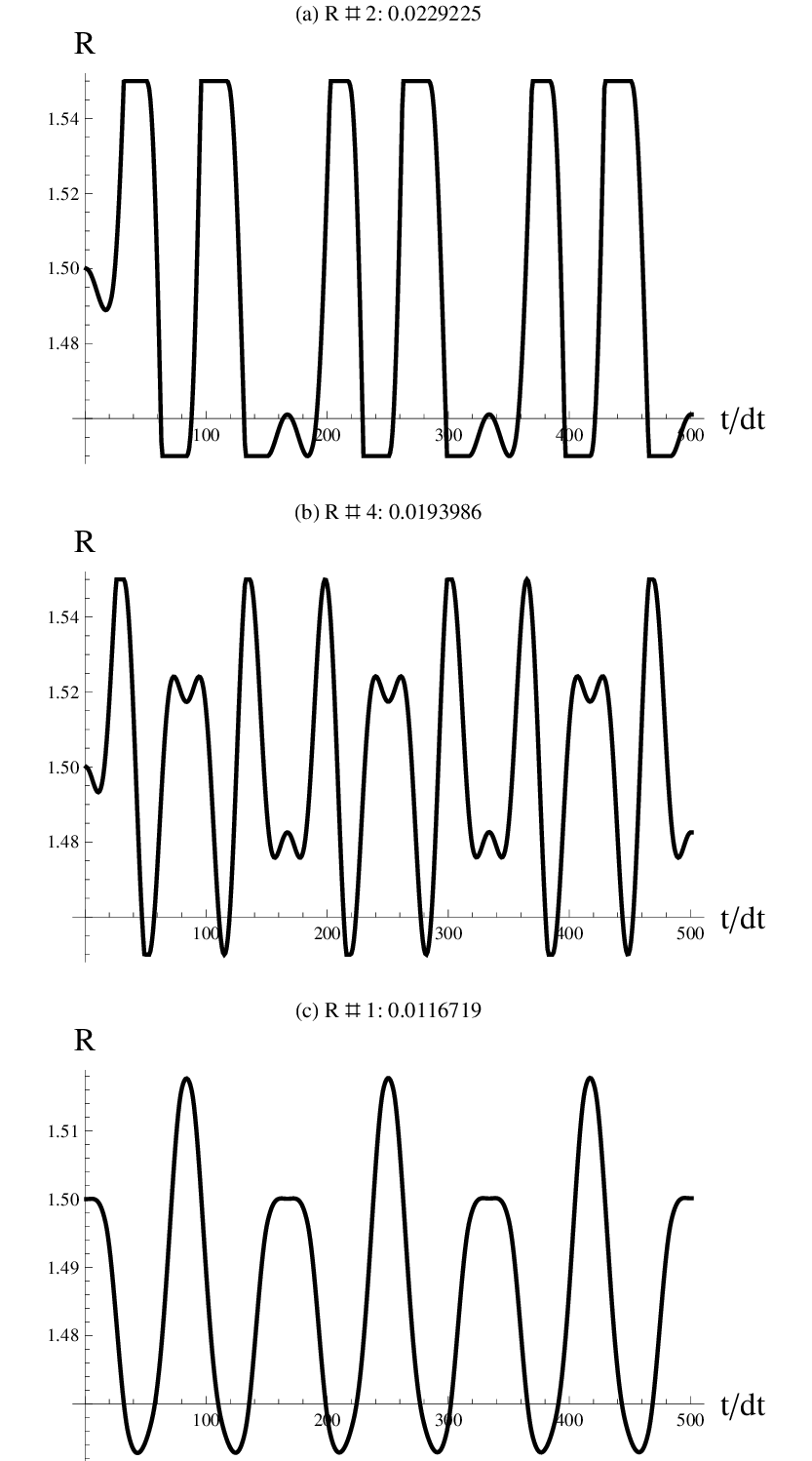}
\end{flushright}
\caption{\label{fig:figCubeTheMostInterestingRM}
The time dependence of the three memristances with the largest dissimilarity measure. They are hardest to mimic by the optimized linear combination of input signals. Note that it would be very hard, if not impossible, to identify them by just inspecting their time dependence.
}
\end{figure}

Figure~\ref{fig:figCubeFourierSpectrumRM} shows the Fourier spectrum \(FS(\omega)\) of the resistances \(R(t)\) of the cube links in Fig.~\ref{fig:figCube}. There is a distinct difference in the appearance of the first two panels (a and b) and the last panel (c). In the first two panels there is part of the spectrum that cannot be obtained by combining input signals in a trivial (linear) way. The signals described by the upper two memristances contain extra information that cannot be extracted from the input by using a simple linear filter.  The large peak at the zero frequency comes from the fact that the memristances always stay within their hard limits.  Since the information contained in the peak is trivial, it can be safely ignored. We believe that the observed effect in the other part of the spectrum is genuine and significant.
Figure \ref{fig:figCubeTheMostInterestingRM} shows the time dependence of the three memristances with the largest dissimilarity measure. They are hardest to mimic by the optimized linear combination of input signals. Note that it would be very hard, if not impossible, to identify them by just inspecting their time dependence.


There are other works dealing with the generation of harmonics in memristor networks, e.g. see \cite{OskoeAndSahimiPRE2011,GeorgiouBarahonaYalirakiDrakakis2012} and references therein. These studies deal with an entirely different set of issues. Our goal is to use the harmonics generation  as a very simple test of the quality of the reservoir. The main hypothesis is that if the nonlinear frequency response function of a network of nonlinear systems cannot be approximated, in general, by a linear mixture of delayed inputs, then the quality of the reservoir is good. The reservoir generates additional dynamic states. Other procedures have been suggested for quantifying the quality of a reservoir. For example, the echo state property (fading memory) is fundamental for the computational performance of the reservoir \cite{ManjunathJaeger2013}.

If our hypothesis is correct, Figs.~\ref{fig:figCubeVextVintRM}-\ref{fig:figCubeTheMostInterestingRM} suggest that resistances could be more useful for information processing purposes than the individual node voltages. From an experimental point of view, the resistances are not easily measured, and that also goes for the currents. Instead, one could measure the time-dependent voltage differences across memristor links (neighbouring nodes) or between more distant nodes. Another option could exploit correlation functions between nodes \(i\) and \(j\), \(\langle V_\text{int,j}(t') V_\text{int,i}(t)\rangle\), or even higher order correlation functions.


\section{About computational capacity}

Questions of computational functionality and capacity of networks of switches and gates as dynamical systems have a long history \cite{Turing1948,Gardner1988,Langton1990,Levina2007,Nykter2008}. 

To mathematically define computational capacity of a device is a highly non-trivial task. In principle such a quantity can be estimated by counting how many functions the device can compute. In this context, computing can carry a rather abstract meaning. For example, pattern recognition could be seen as a mapping from the set of input patterns to the Boolean set. Of particular interest has been to connect computational capacity to critical behaviour and the proximity to phase transitions and chaotic dynamics in complex systems \cite{Natschlager2004,Bertschinger2004,Derrida1986,Derrida1987,Gardner1988,Langton1990}, also in the brain  \cite{Kitzbichler2009}.

A natural idea is then to build networks with nanoscale switching elements based on memristive junctions and investigate their properties as dynamical systems and their capacity for computing \cite{Kulkarni2012,KonkoliWendin2013}. Stieg et al. \cite{Stieg2012} create a random network of silver nanowires which is then functionalised by growing Ag-sulfide memristive junctions.
In the present project \cite{Wendin2012,Beiu2012} the nanofabric is of two kinds: (i) a lithographically defined network of organic transistors (NOMFET) \cite{Alibart2009,Alibart2012}, and (ii) a self-assembled network of gold nanoparticles functionalised with molecular switches  \cite{Liao2010,Mangold2011}. It should be possible to extend this approach by functionalizing nanoparticles to form solid-state solid-ionic memristive junctions. In this context it is of interest to note that Strukov and Likarev  \cite{Strukov2011} are proposing to use memristor technology to create NDR elements for all-NDR digital computation in crossbar structures. A network of such NDR elements would also be of interest for investigating of RC type of computation. 

\section{Outlook}

The concept of RC may be particularly useful for nanoscale unconventional computational substrates. The top-down scaling of CMOS already produces networks of extreme nanoscale transistors for digital architectures and computation. However, reliability is an increasing problem, and it is of paramount importance to develop practical schemes for computing with unreliable components. Concerning the role of molecular electronics for ultimate downscaling, typical visions have involved (i) networks of molecular transistors for digital computation, or (ii) neuromorphic networks based on molecular components. The problem with the first approach is that there are no working digital devices that can compete with CMOS. Neuromorphic networks have greater potential, but are far from biological neural networks because of the simplicity of artificial neurons and the limited network connectivity. 

Here reservoir computing - RC - offers a new robust approach that can make use of dense networks of simple switches with memory that form dynamical systems with interesting critical behaviour. The idea is then to use a dynamical system with memory and dissipation as a mixer of a (large) number of input variables, spread over a range of input ports, input frequencies and input times. This can be regarded as a time-dependent pattern generator, and the task is then to analyse the pattern and characterise in real time essential properties of the input signals and their correlations. There are indications that the brain may partly work along RC principles  \cite{Yamazaki2007}.

{\bf Acknowledgements:} This work has been supported by the EU project  FP7-318597 SYMONE and by Chalmers University of Technology.



\newpage
\section*{Figures}
 

%
%

\begin{figure}[H]
\begin{center}
\includegraphics[width=0.7\columnwidth]{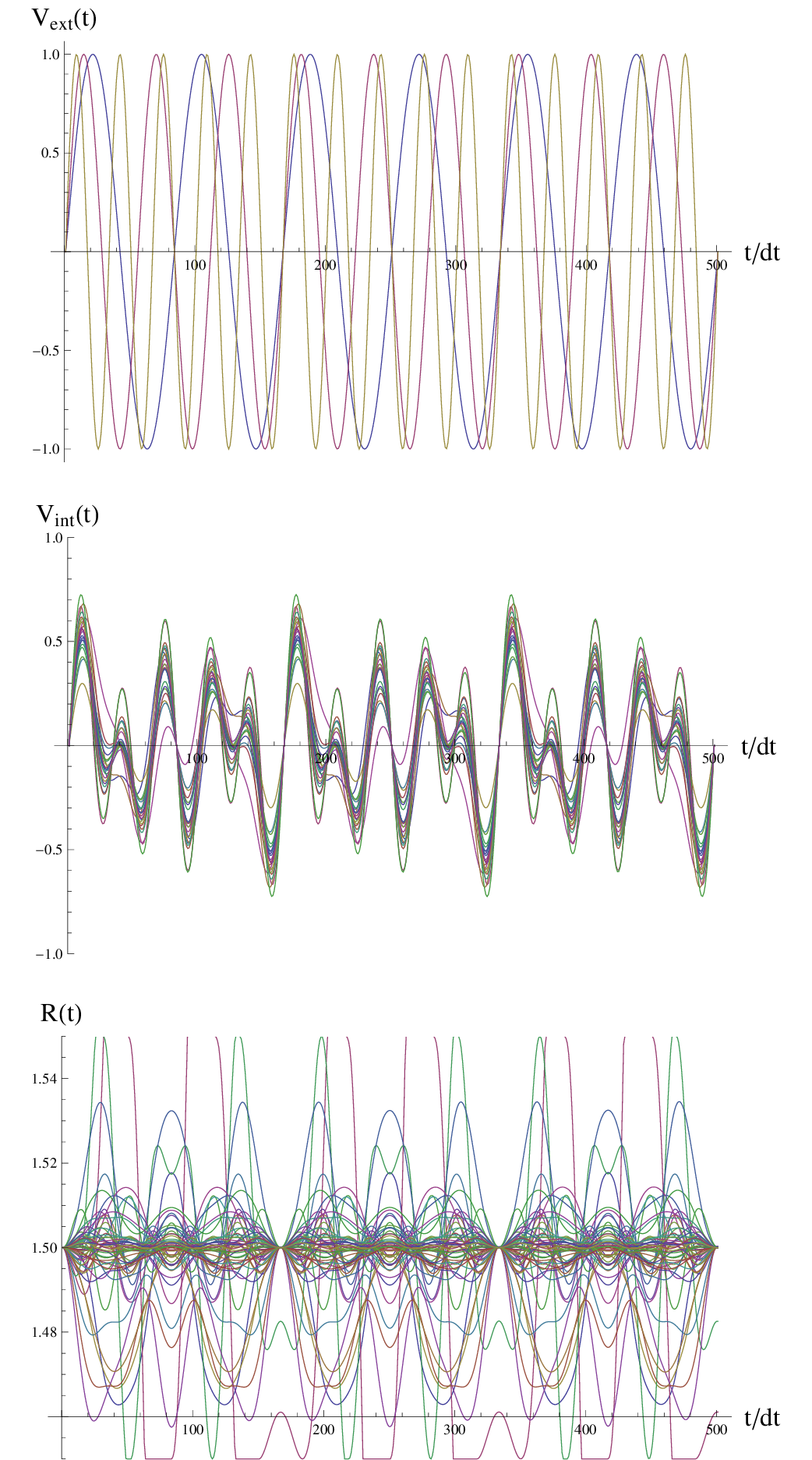}
\end{center}
\caption{\label{fig:figCubeVextVintRM}
Results of simulations for the network in Fig.~\ref{fig:figCube}. The top panel depicts the input voltages, the middle panel depicts the resulting internal voltages, and  the bottom panel depicts the resulting resistances of all memristors.
}
\end{figure}

\begin{figure}[H]
\begin{center}
\includegraphics[width=0.7\columnwidth]{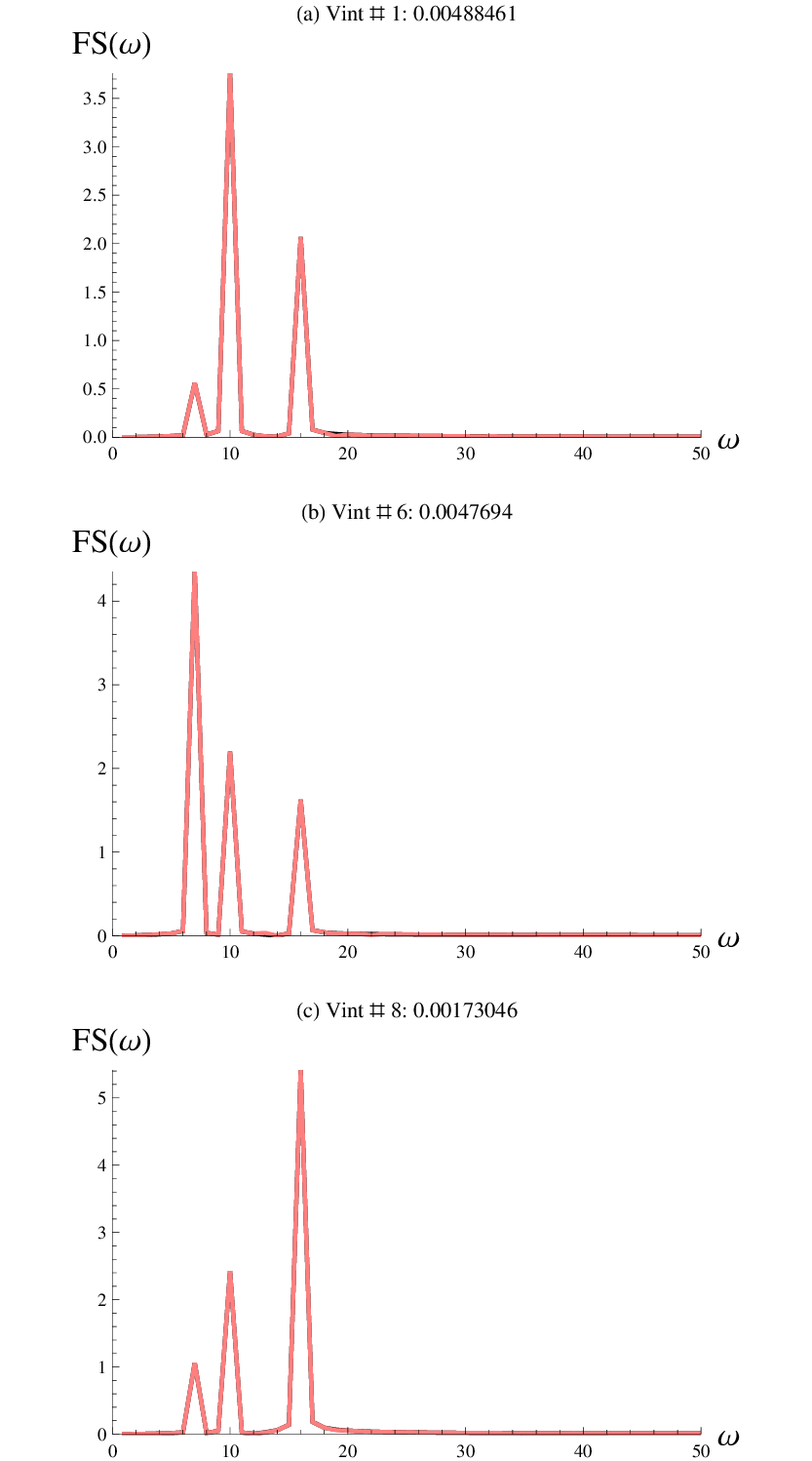}
\end{center}
\caption{\label{fig:figCubeFourierSpectrumVint}
Fourier spectra FS(\(\omega\)) of  the optimised input \(z(t)\) (black) together with output signals \(z(\omega)\) (internal voltages, red) as for Fig.~\ref{fig:figCubeVextVintRM}. The graphs cannot be distinguished visually. The FS(\(\omega\)) is in arbitrary units. The axes labeled by "\(\omega\)" depict integers \(k\) where \(\omega(k) = 2\pi k/(Ndt)\). The first two panels depict the two voltages that are hardest to mimic by the optimized linear combination of input signals (their dissimilarity measures are largest). The last panel depicts the voltage with the smallest dissimilarity measure (that is easiest to mimic by the linear combination of inputs). The similarity measures are very close.
}
\end{figure}

 \begin{figure}[H]
\begin{flushright}
\includegraphics[width=0.7\columnwidth]{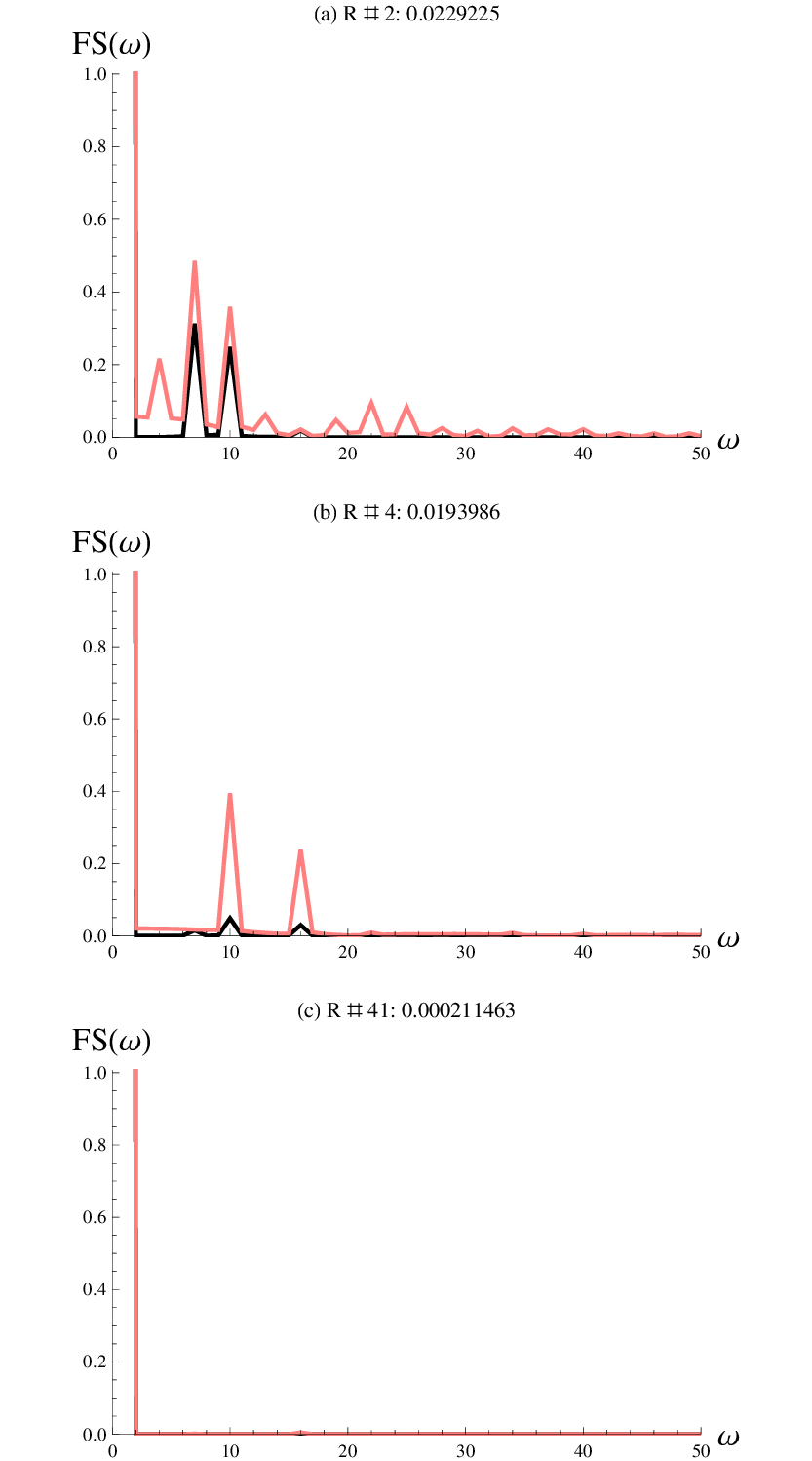}
\end{flushright}
\caption{\label{fig:figCubeFourierSpectrumRM}
Same as figure \ref{fig:figCubeFourierSpectrumVint} but for resistances \(R(t)\). The enumeration of memristors (cube links, Fig.~\ref{fig:figCube}) is unrelated to the node numbers. The first two panels depict the two memristances that are hardest to mimic by the optimized linear combination of input signals (their dissimilarity-measures are the largest). The last panel depicts the memristance that is easiest to mimic.
}
\end{figure}

\newpage

\appendix

\section{A detailed specification of the network} 
\label{app:networkDetails}

The memristor elements of the network depicted in Fig.~\ref{fig:figCubeVextVintRM} are defined as follows. The memristance values are limited to the range \(R_\text{min} = 1.45\Omega\) and \(R_\text{max} = 1.55 \Omega\), and all are initialized at \(R=1.5\Omega\).  The memristor parameters \(\alpha\), \(\beta\), and \(V_T\) for each memristor were chosen at random and their values are given in table~\ref{tab:appendixNetworkDetails}. 

\begin{table}[H]
\caption{\label{tab:appendixNetworkDetails}
The details of the cubic memristor network. Units: \(\alpha\) and \(\beta\) in \(\Omega (Vs)^{-1}\); \(V_T\) in Volts. The direction of a link  \(i\rightarrow j\) is defined as follows. If a voltage \(V_i\) is applied at the node \(i\) and a voltage \(V_j\) at the node \(j\), then the rate of the change of the memristance \(f\) can be obtained from Fig.~\ref{fig:figFuncF} by taking \(V=V_i-V_j\).
} 
\begin{tabular}{@{}*{8}{l}}


\midrule                             
$link$ & $V_T$ & $\alpha$ & $\beta$ & 
   $link$ & $V_T$&$\alpha$& $\beta$ \cr 
\midrule
$10 \rightarrow 1$ & 0.374442 & 0.339527 & 0.76859 & $15 \rightarrow 12$ & 0.459023 & 0.830813 & 1.71065 \cr
$1 \rightarrow 2$ & 0.338867 & 0.766728 & 1.681 & $12 \rightarrow 21$ & 0.248844 & 0.716935 & 0.964279 \cr
$4 \rightarrow 1$ & 0.916852 & 0.190449 & 1.01836 & $13 \rightarrow 22$ & 0.591142 & 0.685118 & 0.850874 \cr
$2 \rightarrow 3$ & 0.395556 & 0.59712 & 0.715665 & $14 \rightarrow 13$ & 0.417162 & 0.503876 & 1.20527 \cr
$2 \rightarrow 5$ & 0.957822 & 0.552519 & 1.38086 & $16 \rightarrow 13$ & 0.330005 & 0.188323 & 1.12526 \cr
$2 \rightarrow 11$ & 0.456982 & 0.197251 & 0.499046 & $14 \rightarrow 15$ & 0.640415 & 0.889584 & 1.77304 \cr
$12 \rightarrow 3$ & 0.509442 & 0.193125 & 0.38095 & $14 \rightarrow 17$ & 0.268105 & 0.826208 & 0.946175 \cr
$3 \rightarrow 6$ & 0.129586 & 0.379953 & 0.580798 & $14 \rightarrow 23$ & 0.976 & 0.920302 & 1.71634 \cr
$4 \rightarrow 7$ & 0.63842 & 0.816642 & 1.21164 & $24 \rightarrow 15$ & 0.124989 & 0.257296 & 0.473974 \cr
$4 \rightarrow 5$ & 0.497659 & 0.89305 & 1.24712 & $15 \rightarrow 18$ & 0.761428 & 0.73645 & 1.17762 \cr
$4 \rightarrow 13$ & 0.812996 & 0.836584 & 1.11225 & $25 \rightarrow 16$ & 0.848135 & 0.475557 & 1.45515 \cr
$5 \rightarrow 8$ & 0.312802 & 0.599889 & 0.799181 & $17 \rightarrow 16$ & 0.134474 & 0.606631 & 1.58427 \cr
$14 \rightarrow 5$ & 0.424215 & 0.973887 & 1.53969 & $26 \rightarrow 17$ & 0.879447 & 0.610327 & 0.764154 \cr
$5 \rightarrow 6$ & 0.211621 & 0.656048 & 0.761047 & $17 \rightarrow 18$ & 0.80575 & 0.205049 & 0.8331 \cr
$9 \rightarrow 6$ & 0.95086 & 0.670889 & 0.970032 & $18 \rightarrow 27$ & 0.164033 & 0.458028 & 1.00478 \cr
$6 \rightarrow 15$ & 0.501306 & 0.433332 & 0.782898 & $20 \rightarrow 19$ & 0.263635 & 0.958362 & 1.59943 \cr
$16 \rightarrow 7$ & 0.746284 & 0.571936 & 1.28032 & $19 \rightarrow 22$ & 0.319153 & 0.679248 & 0.933867 \cr
$7 \rightarrow 8$ & 0.367929 & 0.584094 & 1.23548 & $23 \rightarrow 20$ & 0.374859 & 0.436996 & 0.831076 \cr
$8 \rightarrow 17$ & 0.607931 & 0.824092 & 1.78827 & $21 \rightarrow 20$ & 0.110315 & 0.223772 & 0.589538 \cr
$8 \rightarrow 9$ & 0.955427 & 0.970764 & 1.62366 & $21 \rightarrow 24$ & 0.448737 & 0.352571 & 0.710772 \cr
$9 \rightarrow 18$ & 0.734346 & 0.553811 & 0.963794 & $22 \rightarrow 23$ & 0.803082 & 0.646101 & 0.806519 \cr
$10 \rightarrow 19$ & 0.348097 & 0.603011 & 0.71191 & $25 \rightarrow 22$ & 0.655003 & 0.947564 & 1.39472 \cr
$10 \rightarrow 11$ & 0.63911 & 0.522766 & 0.656083 & $23 \rightarrow 24$ & 0.394366 & 0.693992 & 1.68974 \cr
$13 \rightarrow 10$ & 0.983575 & 0.770146 & 1.51798 & $23 \rightarrow 26$ & 0.790899 & 0.792383 & 1.54045 \cr
$11 \rightarrow 14$ & 0.125702 & 0.702939 & 1.50984 & $24 \rightarrow 27$ & 0.587119 & 0.493326 & 0.967518 \cr
$11 \rightarrow 12$ & 0.292193 & 0.603787 & 1.14385 & $26 \rightarrow 25$ & 0.253639 & 0.787869 & 1.65719 \cr
$20 \rightarrow 11$ & 0.363118 & 0.711326 & 1.61589 & $27 \rightarrow 26$ & 0.388202 & 0.511768 & 0.809962 \cr
\midrule
\end{tabular}
\end{table}

\end{document}